\documentclass[]{tPHM2e}

\begin{document}
\doi{10.1080/14786435.20xx.xxxxxx} \issn{1478-6443}
\issnp{1478-6435} \jvol{00} \jnum{00} \jyear{2010}

\markboth{Fei Han et al.}{Philosophical Magazine}

\articletype{}

\title{Metastable superconducting state in quenched K$_x$Fe$_{2-y}$Se$_2$}

\author{Fei Han$^{\rm a}$, Huan Yang$^{\rm b}$, Bing Shen$^{\rm a}$, Zheng-Yu Wang$^{\rm a}$, Chun-Hong Li$^{\rm a}$ and Hai-Hu Wen$^{\rm b}$$^{\ast}$\thanks{$^\ast$Corresponding author. Email: hhwen@nju.edu.cn\vspace{6pt}}
\\\vspace{6pt}  $^{\rm a}${\em{National Laboratory
for Superconductivity, Institute of Physics and Beijing National
Laboratory for Condensed Matter Physics, Chinese Academy of
Sciences, Beijing 100190, China}}; $^{\rm b}${\em{Center for Superconducting Physics and Materials, National
Laboratory of Solid State Microstructures and Department of
Physics, Nanjing University, Nanjing 210093,
China}}\\\vspace{6pt}\received{v4.5 released May 2010}}

\maketitle

\begin{abstract}
By direct quenching or post-annealing followed by quenching, we
have successfully obtained a series of K$_x$Fe$_{2-y}$Se$_2$
samples with different properties. It is found that the samples
directly quenched in the cooling process of growth show
superconductivity and the one cooled with furnace is insulating
even though their stoichiometries are similar. The sample cooled
with furnace can be tuned from insulating to superconducting by
post-annealing and then quenching. Based on the two points
mentioned above, we conclude that the superconducting state in
K$_x$Fe$_{2-y}$Se$_2$ is metastable, and quenching is the key
point to achieve the superconducting state. The similar
stoichiometries of all the non-superconducting and superconducting
samples indicate that the iron valence doesn't play a decisive
role in determining whether a K$_x$Fe$_{2-y}$Se$_2$ sample is
superconducting. Combining with the result got in the
K$_x$Fe$_{2-y}$Se$_2$ thin films prepared by molecular beam
epitaxy (MBE), we argue that our superconducting sample partly
corresponds to the phase without iron vacancies as evidenced by
scanning tunnelling microscopy (STM) and the insulating sample
mainly corresponds to the phase with the $\sqrt{5}\times\sqrt{5}$
vacancy order. Quenching may play a role of freezing the phase
without iron vacancies.

\begin{keywords}K$_x$Fe$_{2-y}$Se$_2$; direct quenching; post-annealing and then quenching; insulating; superconducting; metastable; iron vacancies
\end{keywords}

\end{abstract}

\section{Introduction}
Since the discovery of superconductivity at 26 K in oxy-pnictide
LaFeAsO$_{1-x}$F$_x$\cite{Hosono}, enormous interests have been
stimulated in the fields of condensed matter physics and material
sciences. Among the several types of iron based superconductors
with different
structures\cite{Hosono,Rotter,ChuCW,WangXC,ChuCW2,WuMK,Cava,VFeAs21311},
FeSe with the PbO structure has received special attention since
its structure is simpler than other iron pnictide
superconductors. However, the superconducting transition
temperature (T$_c$) in iron chalcogenide compounds is not enhanced
as high as other iron pnictide superconductors under ambient
pressure until the superconductivity at above 30 K in K$_x$Fe$_{2-y}$Se$_2$ is
discovered\cite{ChenXL}. The insulating and the superconducting
state are both observed in K$_x$Fe$_{2-y}$Se$_2$ with different
stoichiometries and some groups have tuned the system from
insulating to superconducting by varying the ratio of starting
materials\cite{ChenGF,FangMH,ChenXH}. Here we give two new tuning
methods: direct quenching or post-annealing followed by quenching.
On one hand, by directly quenching at different furnace
temperatures in the cooling process of growth we can get a series
of K$_x$Fe$_{2-y}$Se$_2$ samples with different superconducting
properties, while the sample cooled with furnace slowly is
non-superconducting but insulating. On the other hand, by
post-annealing and then quenching we can tune the previous
insulating K$_x$Fe$_{2-y}$Se$_2$ sample into superconducting state
again, which was discovered by us for the first time and confirmed
by another group\cite{Petrovic}. We also find that the tuning is
reversible, since the superconducting state disappears about 20
days later and the insulating state comes out again in the
post-annealed and quenched crystals. As mentioned above, we think
that the quenching is important to the appearance of
superconductivity and the superconducting state which needs to be
frozen by quenching is metastable.

\section{Sample preparation}
By using the self-flux method, we successfully grown high-quality
single-crystalline samples of K$_x$Fe$_{2-y}$Se$_2$. First FeSe powders were obtained by the
chemical reaction method with Fe powders (purity 99.99\%) and Se
powders(purity 99.99\%). Then the starting materials in the fixed
ratio of K: FeSe = 0.8: 2 were placed in an alumina crucible and
sealed in a quartz tube under vacuum. All the weighing, mixing,
grounding and pressing procedures were finished in a glove box
under argon atmosphere with the moisture and oxygen below 0.1 PPM.
The contents were then heated up to 1030 $^o$C for 3 hours.
Subsequently the furnace was cooled down to 750 $^o$C at a rate of
5 $^o$C/h. Below 750 $^o$C, the sample cooled with furnace was
kept in furnace and cooled down slowly to room temperature while
the directly quenched samples were took out from furnace and
quenched in air at different furnace temperatures. We cleaved some crystals from the previous sample cooled with furnace in the glove box, put them in a one-end-sealed quartz tube, and sealed the other end of the quartz tube with a closed valve. The valve was then connected with pump and opened under vacuum. To protect the crystals from heat, the quartz tube was wrapt with wet paper. All these made the quartz tube sealing procedure be performed with the crystals not exposed to air and not heated. After tube sealing, a post-annealing procedure is carried out on the crystals (enclosed in the evacuated quartz tube) with a heating plate at different temperatures for 1
hour and then the crystals were rapidly removed from the heating stage.

\section{Experimental data and discussion}
The X-ray diffraction (XRD) measurements of our samples were
carried out on a $Mac-Science$ MXP18A-HF equipment with a scanning
range of 10$^\circ$ to 80$^\circ$ and a step of 0.01$^\circ$. The
DC magnetization measurements were done with a superconducting
quantum interference device (Quantum Design, SQUID, MPMS-7T). The
resistance data were collected using a four-probe technique on the
Quantum Design instrument physical property measurement system
(Quantum Design, PPMS-9T) with magnetic fields up to 9$\;$T. The
electric contacts were made using silver paste at room
temperature. The data acquisition was done using a DC mode of the
PPMS, which measures the voltage under an alternative DC current
and the sample resistivity is obtained by averaging these signals
at each temperature. In this way the contacting thermal power is
naturally removed. The temperature stabilization was better than
0.1$\%$ and the resolution of the voltmeter was better than
10$\;$nV.

\subsection{Direct quenching}
\begin{figure}
\includegraphics[width=13cm]{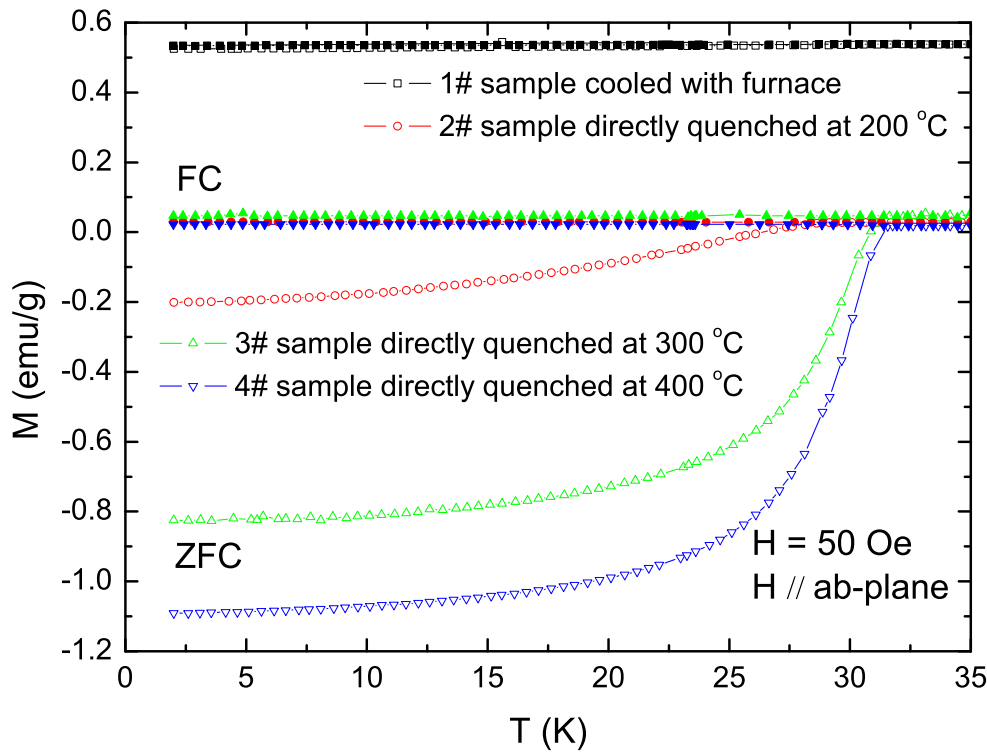}
\caption{(Color online) Temperate dependence of dc magnetization
for the sample cooled with furnace and the samples directly
quenched at about 200 $^o$C, 300 $^o$C, and 400 $^o$C,
respectively. The measurements were carried out under a magnetic
field of 50 Oe in zero-field cooled (ZFC) and field-cooled (FC)
processes with the field parallel to the ab-plane.} \label{fig1}
\end{figure}

In Fig.1 we show the temperate dependence of dc magnetization for
the sample cooled with furnace and the samples directly quenched
at about 200 $^o$C, 300 $^o$C, and 400 $^o$C, respectively. The
measurements were carried out under a magnetic field of 50 Oe in
zero-field cooled (ZFC) and field-cooled (FC) processes with the
field parallel to the ab-plane. Paramagnetic signal is observed in
the sample cooled with furnace and there is not diamagnetization
in the low temperature regime. A weak diamagnetic signal, which is
corresponding to superconductivity, appears below about 28 K in
the sample directly quenched at 200 $^o$C. When the quenching
temperature increases to above 300 $^o$C, strong diamagnetic
signals appear below about 31.5 K. We find that the quenching
temperature has an important influence on the diamagnetization
signal. In sharp contrast to it, we find that the transition
temperature does not strongly depend on the quenching temperature
since the diamagnetization signals all appear below 28-32 K in the
samples directly quenched at 200 $^o$C, 300 $^o$C, and 400 $^o$C.

\begin{figure}
\includegraphics[width=13cm]{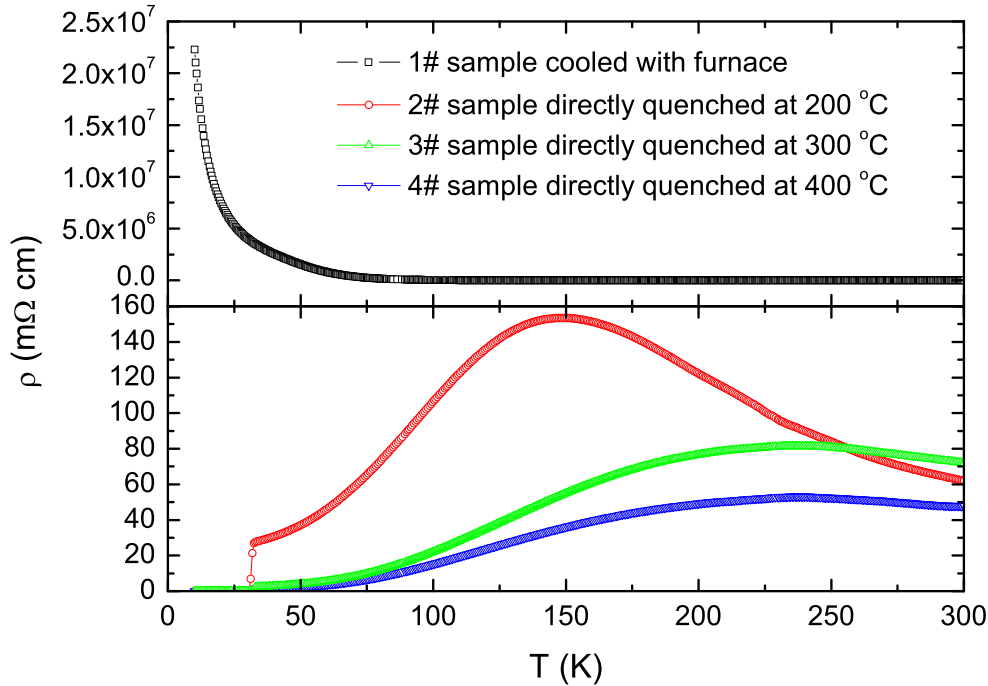}
\caption{(Color online) Temperature dependence of resistivity for
the sample cooled with furnace and the samples directly quenched
at about 200 $^o$C, 300 $^o$C, and 400 $^o$C, respectively.}
\label{fig1}
\end{figure}

By electrical resistivity measurements we find that the
non-superconducting sample cooled with furnace has a insulating
behavior, as shown in in the top panel of Fig.2. The sample
directly quenched at 200 $^o$C has a superconducting transition at
32.5 K and a hump-like anomaly at 150 K in the curve of $\rho$(T).
The sample directly quenched at 300 $^o$C and 400 $^o$C is also
superconducting with the same T$_c$ as the sample directly
quenched at 200 $^o$C and the hump-like anomaly shifts to about
250 K. We find that the absolute value of resistivity decreases
with the quenching temperature increasing from 200 to 400 $^o$C,
which is a strong support to that K$_x$Fe$_{2-y}$Se$_2$ is a
phase-separation system composed of a metallic phase and a
insulating phase.

We perform magnetization and resistivity measurements for several times and find both the magnetization and the resistivity results are reproducible, which indicates that the insulating property in the sample cooled with furnace and the superconductivity in the directly quenched samples is bulk.

To investigate what effect the quenching procedure has on the
K$_x$Fe$_{2-y}$Se$_2$ samples and why the superconductivity
appears, we carried out X-ray diffractions on these samples and
used inductively coupled plasma (ICP) to determine the
stoichiometries of K$_x$Fe$_{2-y}$Se$_2$ samples.

\begin{figure}
\includegraphics[width=13cm]{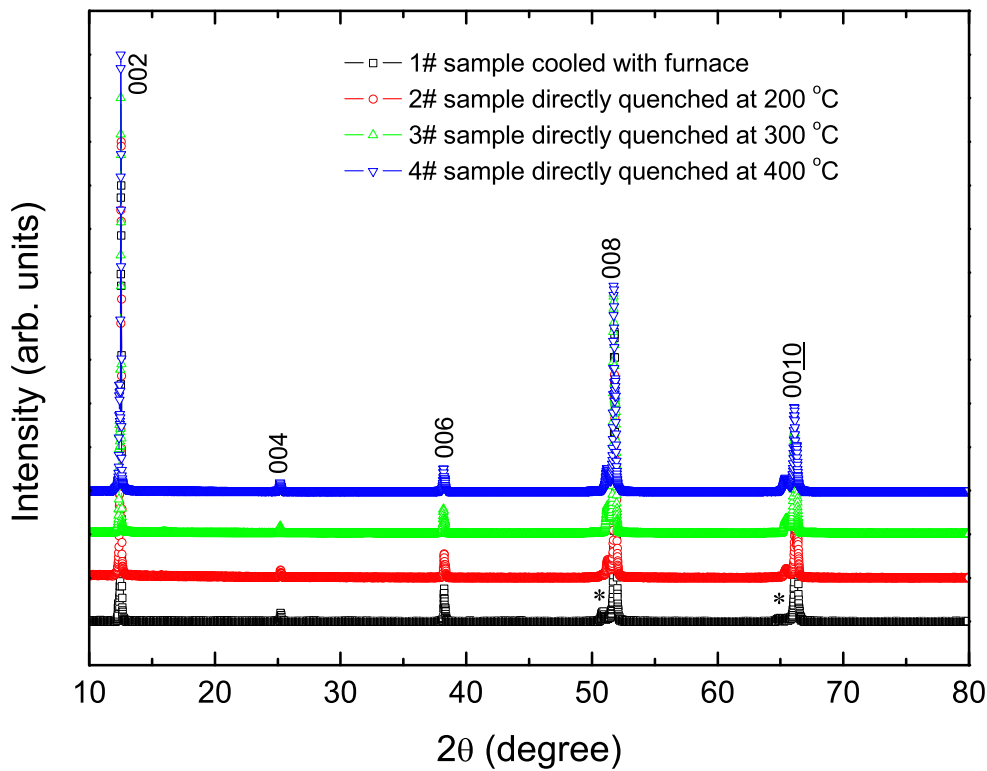}
\caption{(Color online) X-ray diffraction patterns showing the
(00$l$) reflections from the basal plane of the
K$_x$Fe$_{2-y}$Se$_2$ sample cooled with furnace and the samples
directly quenched at 200 $^o$C, 300 $^o$C, and 400 $^o$C.}
\label{fig1}
\end{figure}

As shown in Fig.3, the peaks from the (00$l$) reflections are
still very sharp, indicating excellent crystalline quality. And we
hardly find very obvious shifting among these peaks. However, the
peaks marked by the asterisks in the samples directly quenched
seem to shift closer to the nearby (008) and (00$\overline{10}$)
peaks than in the sample cooled with furnace, which possibly
indicates the two weak peaks come from a super-lattice of iron
vacancies order and the phase with iron vacancies order peters out
after quenching.

\begin{table}[!h]
\tabcolsep 0pt \caption{Stoichiometries and iron valences for the
sample cooled with furnace and the samples directly quenched at
different temperatures while the nominal stoichiometries of these
samples are fixed as K$_{0.8}$Fe$_{2}$Se$_{2}$.} \vspace*{6pt}
\begin{center}
\def\temptablewidth{1\textwidth}
{\rule{\temptablewidth}{1pt}}
\begin{tabular*}{\temptablewidth}{@{\extracolsep{\fill}}ccccccc}
Quenching temperature  & Stoichiometry  & Valence of
iron\\
\hline
       Cooled with furnace &K$_{0.80}$Fe$_{1.69}$Se$_{2}$ &1.890 \\
       200 $^o$C &K$_{0.76}$Fe$_{1.71}$Se$_{2}$ &1.895 \\
       300 $^o$C &K$_{0.78}$Fe$_{1.70}$Se$_{2}$ &1.894 \\
       400 $^o$C &K$_{0.76}$Fe$_{1.70}$Se$_{2}$ &1.906 \\
\end{tabular*}
{\rule{\temptablewidth}{1pt}}
\end{center}
\end{table}

We find that the actual compositions and the iron valences of all
the four samples are very similar to each other. It is noteworthy
that the iron valences are all located at the
non-superconducting region in the electronic and magnetic phase
diagram of K$_x$Fe$_{2-y}$Se$_2$ as a function of iron
valence\cite{ChenXH}. Based on this point, we think that the phase
diagram as a function of iron valence didn't solve the problem what determines K$_x$Fe$_{2-y}$Se$_2$ to be superconducting or not. There must be a more
essential factor working effect instead of iron valence. We notice that different groups including our group got different results even though they prepared samples in a same nominal stoichiometry. When the nominal stoichiometry was K$_{0.8}$Fe$_2$Se$_2$, some groups saw clear superconductivity in the samples not quenched from high temperatures with the actual compositions as K$_{0.78}$Fe$_{1.70}$Se$_2$\cite{ChenXL} and K$_{0.80}$Fe$_{1.76}$Se$_2$\cite{HuRW,HuRW3} respectively. The iron valences of the two samples were also located at the
non-superconducting region in the electronic and magnetic phase
diagram. The phase diagram makers reported that their K$_{0.8}$Fe$_2$Se$_2$ sample was also superconducting and had an actual composition of K$_{0.73}$Fe$_{1.67}$Se$_2$. In sharp contrast to their results, we never see superconductivity in not quenched samples when the nominal composition is K$_{0.8}$Fe$_2$Se$_2$ as well and the actual composition is similar to one of the actual compositions reported by other groups, K$_{0.78}$Fe$_{1.70}$Se$_2$. In consideration of these different results, we think that the property of K$_x$Fe$_{2-y}$Se$_2$ is sensitive to preparation conditions. Many conditions such as the quality of quartz tube and the warm-keeping performance of furnace all have influences on the property of K$_x$Fe$_{2-y}$Se$_2$. For instance, the different actual compositions of K$_x$Fe$_{2-y}$Se$_2$ among the samples prepared by different groups may be caused by the different qualities of quartz tubes which make different loss of K in amount. The different transport properties of K$_x$Fe$_{2-y}$Se$_2$ even among the samples not quenched with similar actual compositions may be caused by the different warm-keeping performances of furnaces which make different cooling rates after the furnaces are turned off. We control the preparation conditions consistent, so our result is repeatable, credible and not contradictory to other different experimental results.

\subsection{Post-annealing and then quenching}
\begin{figure}
\includegraphics[width=13cm]{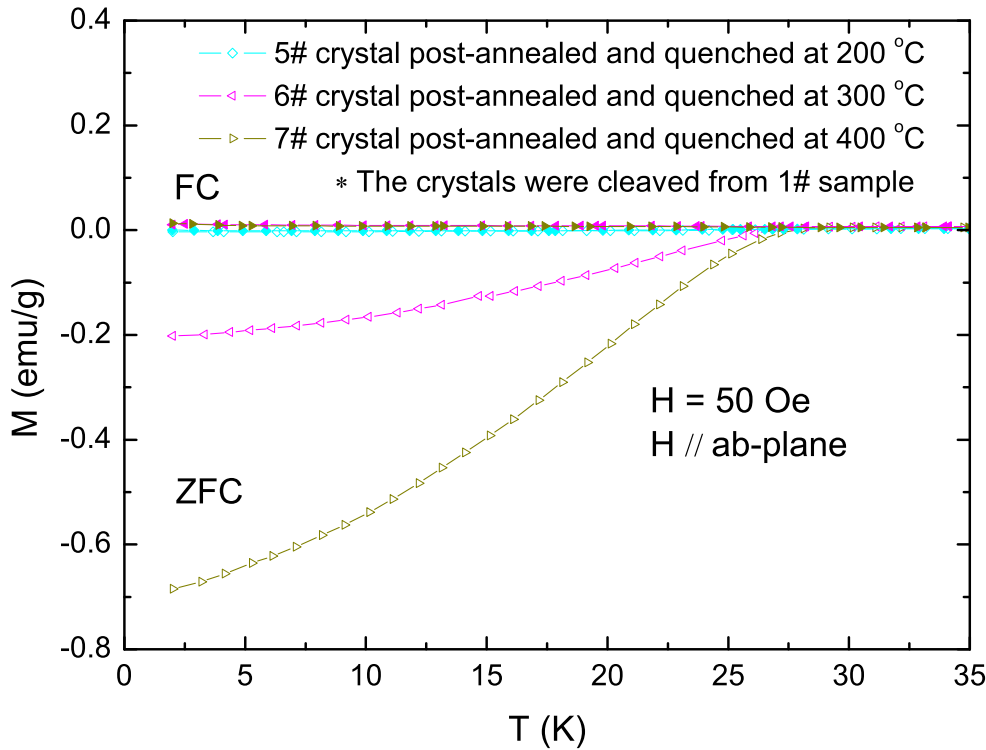}
\caption{(Color online) Temperate dependence of dc magnetization
for the crystals post-annealed and then quenched at at about 200
$^o$C, 300 $^o$C, and 400 $^o$C, which were cleaved from No.1
sample in advance. The measurements were carried out under a
magnetic field of 50 Oe in zero-field cooled (ZFC) and
field-cooled (FC) processes with the field parallel to the
ab-plane.} \label{fig1}
\end{figure}

By post-annealing and then quenching at 200 $^o$C, 300 $^o$C, and
400 $^o$C, we tune the No.1 sample from insulating to
superconducting. In Fig.4 we show the temperate dependence of dc
magnetization for the crystals post-annealed and then quenched at
about 200 $^o$C, 300 $^o$C, and 400 $^o$C, respectively. The
measurements were carried out under a magnetic field of 50 Oe in
zero-fieldcooled (ZFC) and field-cooled (FC) processes with the
field parallel to the ab-plane. There is not a diamagnetic signal
observed in the crystal post-annealed and quenched at 200 $^o$C.
When the annealing and quenching temperature increases to above
300 $^o$C, diamagnetic signals begin to appear below about 26 K.
We find that the annealing and quenching temperature has an
important influence on the diamagnetization signal like the
condition of direct quenching. However, the diamagnetic signals of
the crystals post-annealed and then quenched are not as strong as
the ones of the samples directly quenched under the same quenching
temperatures.

\begin{figure}
\includegraphics[width=13cm]{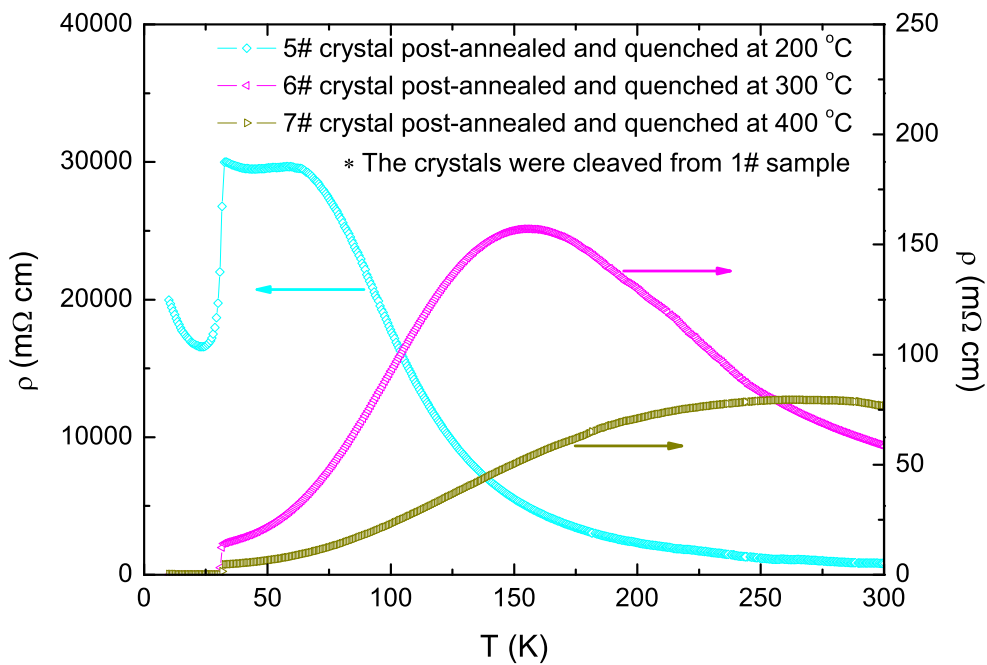}
\caption{(Color online) Temperature dependence of resistivity for
the crystals post-annealed and then quenched at at about 200
$^o$C, 300 $^o$C, and 400 $^o$C, which were cleaved from No.1
sample in advance.} \label{fig1}
\end{figure}

By electrical resistivity measurements we find that the
post-annealed and quenched crystals' transport characters are
getting different from the No.1 sample, as shown in Fig.5. The
crystal after post-annealing and quenching at 200 $^o$C has a
semiconducting/insulating behavior from 30 to 300 K and the
resistivity drops when the temperature decreases below 30 K. The
dropping may imply that more metallic phase is achieved
corresponding to the superconductivity. The crystal post-annealed
and quenched at 300 $^o$C has been tuned to superconducting state
and there is a hump-like anomaly at 155 K in the curve of
$\rho$(T). The crystal post-annealed and quenched at 400 $^o$C is
also superconducting and the hump-like anomaly shifts to about 250
K. We also find that the absolute value of resistivity decreases
with increasing the quenching temperature from 200 to 400 $^o$C.

\begin{figure}
\includegraphics[width=13cm]{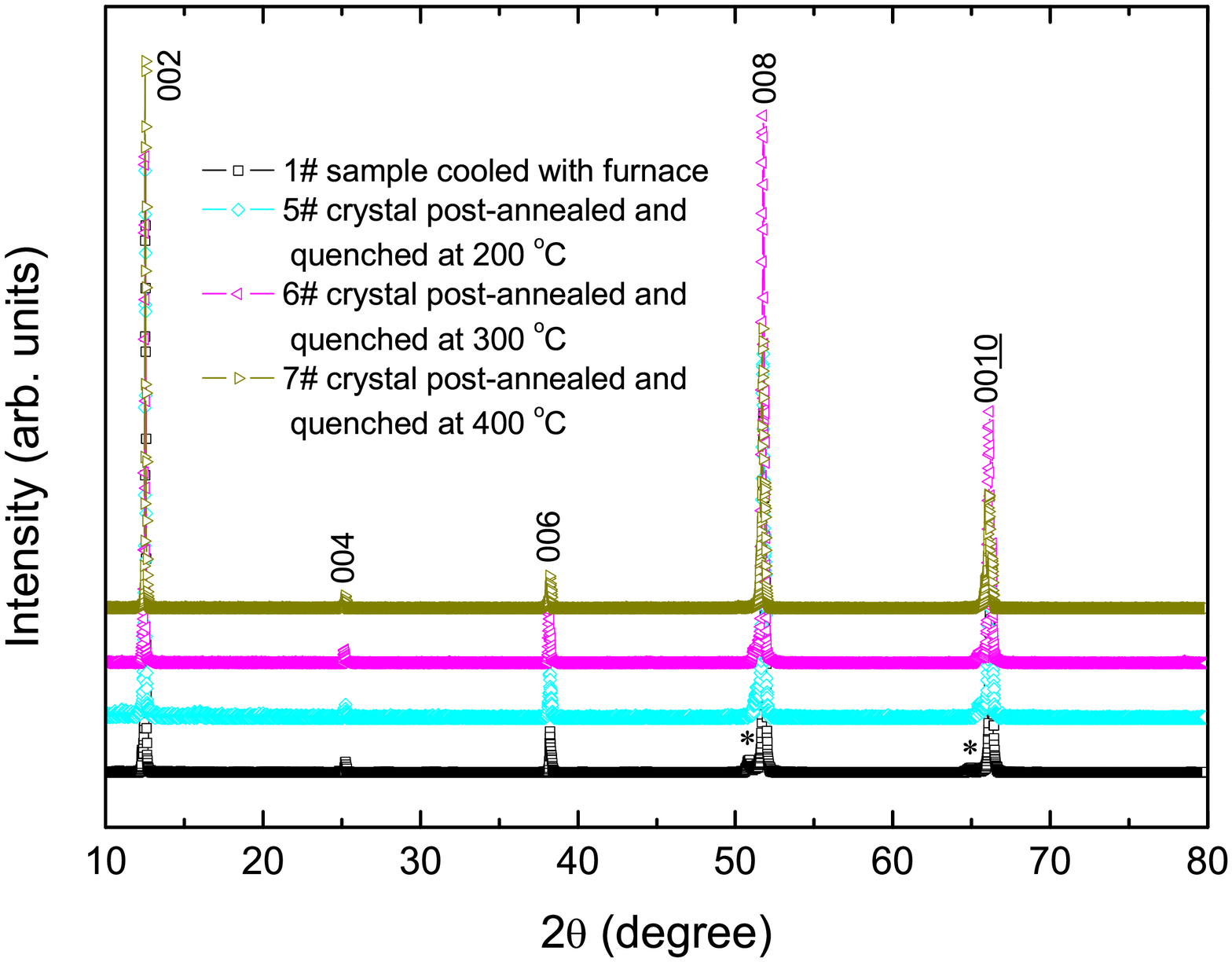}
\caption{(Color online) X-ray diffraction patterns showing the
(00$l$) reflections from the basal plane of the
K$_x$Fe$_{2-y}$Se$_2$ sample cooled with furnace and the crystals
post-annealed and then quenched at about 200 $^o$C, 300 $^o$C, and
400 $^o$C. The samples undergoing different treating processes
were cleaved from the sample cooled with furnace and not
superconducting.} \label{fig1}
\end{figure}

We also carried out X-ray diffraction on these crystals. As shown
in Fig.6, the peaks from the (00$l$) reflections are still very
sharp after annealing. We still hardly find very obvious shifting
among these peaks. And the peaks marked by the asterisks seem to
shift closer to the nearby (008) and (00$\overline{10}$) peaks
after annealing and quenching. This feature is similar to the case
of direct quenching.

\begin{figure}
\includegraphics[width=13cm]{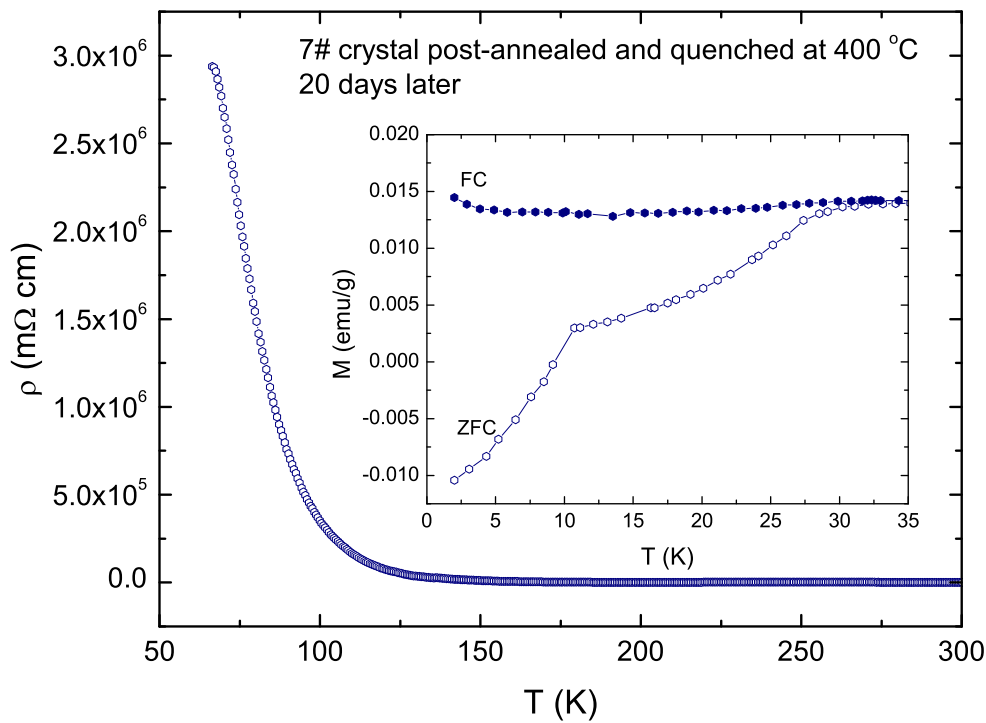}
\caption{(Color online) After 20 days, the temperate dependence of
dc magnetization and resistivity for the crystal post-annealed and
quenched at 400 $^o$C. } \label{fig1}
\end{figure}

Surprisingly, the crystals post-annealed and quenched lost their
superconducting characters after a period of time, for example, 20
days later. In this period the crystals were always kept in the
argon atmosphere. As shown in Fig.7, after 20 days, the strong
diamagnetization signal has disappeared and the insulating state
comes out again in the No.7 crystal. Obviously, the
superconducting state tuned from the insulating state by
post-annealing and quenching is unstable. However, there is no time dependence of superconductivity observed in our directly quenched samples and reported by other groups, which suggests that the freezing effect of post-annealing and quenching is temporary and less effective than directly quenching.

\section{Conclusions}

In summary, we find that the samples directly quenched in the
cooling process of growth show superconducting while the one
cooled with furnace is insulating, and the latter can be tuned
from insulating to superconducting by post-annealing and then
quenching. In addition, the actual compositions and the iron
valences of all the non-superconducting and superconducting
samples are very similar to each other since we fixed the nominal
stoichiometries in preparing process. Based on the two factors, we
conclude that the superconducting state in K$_x$Fe$_{2-y}$Se$_2$
is metastable, and quenching is the key point to achieve the
superconducting state. The similar stoichiometries of all the
non-superconducting and superconducting samples also indicate that
the iron valence doesn't play a decisive role in determining
whether a K$_x$Fe$_{2-y}$Se$_2$ sample is superconducting. Our
result in resistivity indicates that K$_x$Fe$_{2-y}$Se$_2$ is a
phase-separation system composed of a metallic phase and a
insulating phase. Our XRD results suggest that there is a
super-lattice of iron vacancies order in this system and the phase
with iron vacancies is less in the superconducting samples than in
the insulating samples. All these results give a support to the M\"{o}ssbauer result reported before that the superconductivity in K$_x$Fe$_{2-y}$Se$_2$ comes from a minority phase which does not have large moment and the long range magnetic order belongs to a non-superconducting majority phase\cite{HuRW2}. Combining with the result obtained in the K$_x$Fe$_{2-y}$Se$_2$ thin films prepared by molecular beam
epitaxy (MBE)\cite{XueQK}, we argue that our superconducting
sample partly corresponds to the phase without iron vacancies as
seen by scanning tunneling microscopy (STM), and the insulating
sample mainly corresponds to the phase with
$\sqrt{5}\times\sqrt{5}$ iron vacancy order. Quenching may play a
role of freezing the phase without iron vacancies. Therefore,
whether the K$_x$Fe$_{2-y}$Se$_2$ sample contains the phase
without iron vacancies is more essential to achieve
superconductivity, instead, the iron valence does not play an
important role. By varying the ratio of starting materials to tune
the iron valence or quenching a sample with fixed stoichiometry
are both effective to obtain the superconducting state. There is,
perhaps a difference, that the superconducting state tuned by
varying the ratio of starting materials is more stable, while the
superconducting state frozen by quenching is metastable.

\section*{Acknowledgements}
This work is supported by the Natural Science Foundation of China,
the Ministry of Science and Technology of China (973 project:
2011CBA00102).


\begin{thebibliography}{10}

\bibitem{Hosono}Y. Kamihara, T. Watanabe, M. Hirano, and H. Hosono, J. Am. Chem. Soc. 130 (2008) p.3296.

\bibitem{Rotter}M. Rotter, M. Tegel, and D. Johrendt, Phys. Rev. Lett. 101 (2008) p.107006.

\bibitem{ChuCW}K. Sasmal, B. Lv, B. Lorenz, A. Guloy, F. Chen, Y. Xue, and C. W. Chu, Phys. Rev. Lett. 101 (2008) p.107007.

\bibitem{WangXC}X. C. Wang, Q. Q. Liu, Y. X. Lv, W. B. Gao, L. X. Yang, R. C. Yu, F. Y. Li, and C. Q. Jin, Solid State Commun. 148 (2008) p.538.

\bibitem{ChuCW2}J. H. Tapp, Z. Tang, B. Lv, K. Sasmal, B. Lorenz, P. C. W. Chu, and A. M. Guloy, Phys. Rev. B 78 (2008) p.060505(R).

\bibitem{WuMK}F. C. Hsu, J. Y. Luo, K. W. Yeh, T. K. Chen, T. W. Huang, P. M. Wu, Y. C. Lee, Y. L. Huang, Y. Y. Chu, D. C. Yan, and M. K. Wu, Proc. Natl. Acad. Sci. 105 (2008) p.14262.

\bibitem{Cava}T. Klimczuk, T.M. McQueen, A.J. Williams, Q. Huang, F. Ronning, E.D. Bauer, J.D. Thompson, M.A. Green, and R.J. Cava, Phys. Rev. B 79 (2009) p.012505.

\bibitem{VFeAs21311}X. Zhu, F. Han, G. Mu, P. Cheng, B. Shen, B. Zeng, and H. H. Wen, Phys. Rev. B 79 (2009) p.220512(R).

\bibitem{ChenXL}J. Guo, S. Jin, G. Wang, S. Wang, K. Zhu, T. Zhou, M. He and X. Chen, Phys. Rev. B 82 (2010) p.180520(R).

\bibitem{ChenGF}D. M. Wang, J. B. He, T. L. Xia, and G. F. Chen, Phys. Rev. B 83 (2011) p.132502.

\bibitem{FangMH}M. H. Fang, H. D. Wang, C. H. Dong, Z. J. Li, C. M. Feng, J. Chen, and H. Q. Yuan, Europhys. Lett. 94 (2011) p.27009.

\bibitem{ChenXH}Y. J. Yan, M. Zhang, A. F. Wang, J. J. Ying, Z. Y. Li, W. Qin, X. G. Luo, J. Q. Li, J. Hu, and X. H. Chen, arXiv:cond-mat/1104.4941 (2011).

\bibitem{HuRW}R. Hu, K. Cho, H. Kim, H. Hodovanets, W. E. Straszheim, M. A. Tanatar, R. Prozorov, S. L. Bud'ko and P. C. Canfield, Supercond. Sci. Technol. 24 (2011) p.065006.

\bibitem{HuRW3}E. D. Mun, M. M. Altarawneh, C. H. Mielke, V. S. Zapf, R. Hu, S. L. Bud'ko, and P. C. Canfield, Phys. Rev. B 83 (2011) p.100514(R).

\bibitem{Petrovic}H. Lei, and C. Petrovic, arXiv:cond-mat/1110.5316 (2011).

\bibitem{HuRW2}D. H. Ryan, W. N. Rowan-Weetaluktuk, J. M. Cadogan, R. Hu, W. E. Straszheim, S. L. Bud'ko, and P. C. Canfield, Phys. Rev. B 83 (2011) p.104526.

\bibitem{XueQK}W. Li, H. Ding, P. Deng, K. Chang, C. Song, K. He, L. Wang, X. Ma, J. P. Hu, X. Chen, and Q. K. Xue, Nat. Phys. 8 (2012) p.126.

\end{thebibliography}
\end{document}